\documentclass[aps,prl,twocolumn,amsmath,amssymb,superscriptaddress, floatfix]{revtex4}
\usepackage{graphicx}
\usepackage{dcolumn}
\usepackage{bm}
\usepackage{amsmath}
\usepackage{color}
\begin{document}

\title{Synthesis of Homogeneous Manganese-Doped Titanium Oxide Nanotubes from Titanate Precursors}

\author{P\'{e}ter~Szirmai}
\altaffiliation[Present address: ]{Budapest University of Technology and Economics, Institute
of Physics and Condensed Matter Research Group of the Hungarian Academy of
Sciences, H-1521 Budapest, Hungary}
\affiliation{Laboratory of Physics of Complex Matter, \'{E}cole Polytechnique F\'{e}d\'{e}rale de Lausanne, 1015 Lausanne, Switzerland}
\author{Endre~Horv\'{a}th}
\affiliation{Laboratory of Physics of Complex Matter, \'{E}cole Polytechnique F\'{e}d\'{e}rale de Lausanne, 1015 Lausanne, Switzerland}

\author{B\'{a}lint~N\'{a}fr\'{a}di}
\affiliation{Laboratory of Physics of Complex Matter, \'{E}cole Polytechnique F\'{e}d\'{e}rale de Lausanne, 1015 Lausanne, Switzerland}

\author{Zlatko~Mickovi\'{c}}
\affiliation{Laboratory of Physics of Complex Matter, \'{E}cole Polytechnique F\'{e}d\'{e}rale de Lausanne, 1015 Lausanne, Switzerland}

\author{Rita~Smajda}
\affiliation{Laboratory of Physics of Complex Matter, \'{E}cole Polytechnique F\'{e}d\'{e}rale de Lausanne, 1015 Lausanne, Switzerland}
\author{Dejan~M.~Djoki\'{c}}
\affiliation{Laboratory of Physics of Complex Matter, \'{E}cole Polytechnique F\'{e}d\'{e}rale de Lausanne, 1015 Lausanne, Switzerland}
\author{Kurt~Schenk}
\affiliation{Institute of Physics of Biological Systems, \'{E}cole Polytechnique F\'{e}d\'{e}rale de Lausanne, 1015 Lausanne, Switzerland}
\author{L\'{a}szl\'{o} Forr\'{o}}
\affiliation{Laboratory of Physics of Complex Matter, \'{E}cole Polytechnique F\'{e}d\'{e}rale de Lausanne, 1015 Lausanne, Switzerland}
\author{Arnaud~Magrez}
\email{arnaud.magrez@epfl.ch}
\affiliation{Laboratory of Physics of Complex Matter, \'{E}cole Polytechnique F\'{e}d\'{e}rale de Lausanne, 1015 Lausanne, Switzerland}

\begin{abstract}
We report a novel synthesis route of homogeneously manganese-doped TiO$_2$ nanotubes in a broad concentration range. The scroll-type trititanate (H$_2$Ti$_3$O$_7$) nanotubes prepared by hydrothermal synthesis were used as precursors. Mn$^{2+}$ ions were introduced by an ion exchange method resulting Mn$_x$H$_{2-x}$Ti$_3$O$_7$. In a subsequent heat-treatment they were transformed into Mn$_y$Ti$_{1-y}$O$_2$ where $y=x/(3+x)$. The state and the local environment of the Mn$^{2+}$ ions in the precursor and final products were studied by Electron Spin Resonance (ESR) technique. It was found that the Mn$^{2+}$ ions occupy two positions: the first having an almost perfect cubic symmetry while the other is in a strongly distorted octahedral site. The ratio of the two Mn$^{2+}$ sites is independent of the doping level and amounts to 15:85 in Mn$_x$H$_{2-x}$Ti$_3$O$_7$ and to 5:95 in Mn$_y$Ti$_{1-y}$O$_2$. SQUID magnetometry does not show long-range magnetic order in the homogeneously Mn$^{2+}$-doped nanotubes.
\end{abstract}
\maketitle

\section{Introduction}
Titanium dioxide has been extensively studied due to its high thermal and chemical stability, abundance and environmental friendliness. This material is widely used in heterogeneous catalysis and photocatalysis, as white pigment in paints, food and cosmetic products, corrosion-protective coatings and biocompatible layer of bone implants\cite{diebold2003}. Nanostructured TiO$_2$ films are used as photoanode in solar-to-electric energy conversion devices such as dye-sensitized solar cells (DSSCs) \cite{oregan}, in gas sensors \cite{garzella} and in supercapacitors \cite{fabregat}. It is expected that a detailed study of the structural and electronic properties of TiO$_2$ will help in the understanding of the material´s behavior as well as to improve the performances of the previously mentioned applications \cite{forro1993, jacim_epl}.\\
Recently, TiO$_2$ nanotubes and nanowires have received a great deal of attention. These elongated structures possess large surface area, and can be used to prepare novel 3D and highly crystalline structures exhibiting large porosity from which efficient DSSCs are built \cite{tetreault}. Titanate nanowires can serve as a scaffold for self-organization of organic molecules. Based on this property, high sensitivity optical humidity sensors with fast response time have been realized \cite{horvath2012}. Furthermore, TiO$_2$ is a popular material in spintronics \cite{ogale2010}. It is expected to show room temperature ferromagnetism when doped with transition metal ions \cite{matsumoto2001, coey2005}. TiO$_2$ thin films have shown this effect above 5 \% Mn substitutional doping \cite{xu2009, sharma2011}. A similar phenomenon has been observed at low doping level in bulk manganese-doped TiO$_2$ produced by sintering \cite{tian2008}. Ferromagnetism is explained on the basis of the bound magnetic polaron model. However, these results are the subject of controversy as ferromagnetism could arise from impurities, aggregation of doping or magnetic clusters \cite{coey2006}. These flaws could be the product of inhomogeneously prepared TiO$_2$ precursors or the result of segregation caused by the high-temperature synthesis process \cite{nakajima}. This ambiguity underscores the need to elaborate a reliable synthesis method for homogeneous doping of TiO$_2$ with transition metal ions.\\
Here we report a low-temperature synthesis route of homogeneously doped TiO$_2$ nanotubes (NTs) with Mn$^{2+}$ ions using scroll-type trititanate (Na$_2$Ti$_3$O$_7$) nanotube precursor (Na-NTs) produced by an alkali hydrothermal treatment of TiO$_2$. These multiwalled nanotubes are composed of stepped or corrugated host layers of edge-sharing TiO$_6$ octahedrons having interlayer alkali metal cations. By ionic exchange, the alkali titanates can be easily modified into Mn$_x$H$_{2-x}$Ti$_3$O$_7$, a transition-metal-doped protonated titanate (MnH-NTs) with maximal concentration of about $x\approx 0.18$. A subsequent heat treatment transforms it into Mn$_y$Ti$_{1-y}$O$_2$ (Mn-NTs) with $y=x/(x+3)$. The concentration $y$ reaches a maximum of 5.6 at.\%. Here we focus on the spatial distribution of Mn$^{2+}$ in nanotubular titanates and on their magnetic response by using X-ray diffraction (XRD), Electron Spin Resonance spectroscopy (ESR) and SQUID techniques. We found two Mn$^{2+}$ positions in the structure with cubic and strongly distorted octahedral local symmetries. ESR and SQUID measurements do not show a long range magnetic order.

\section{Experimental methods}
In a typical synthesis, 1~g of Titanium (IV) oxide nanopowder (99.7 \% anatase, Aldrich) is mixed with 30~ml of 10~M NaOH (97 \% Aldrich) solution. The mixture is then transferred to a Teflon-lined stainless steel autoclave (Parr Instrument Co.) and heated to $130~^{\circ}$C and kept at this temperature for 72~h. After the treatment, the autoclave is cooled down to room temperature at a rate of $1~^{\circ}$C/min. The obtained Na$_2$Ti$_3$O$_7$ product is then filtered and washed several times with deionized
\begin{figure}[!tb]
\includegraphics*[width=0.9\linewidth,height=0.73\linewidth]{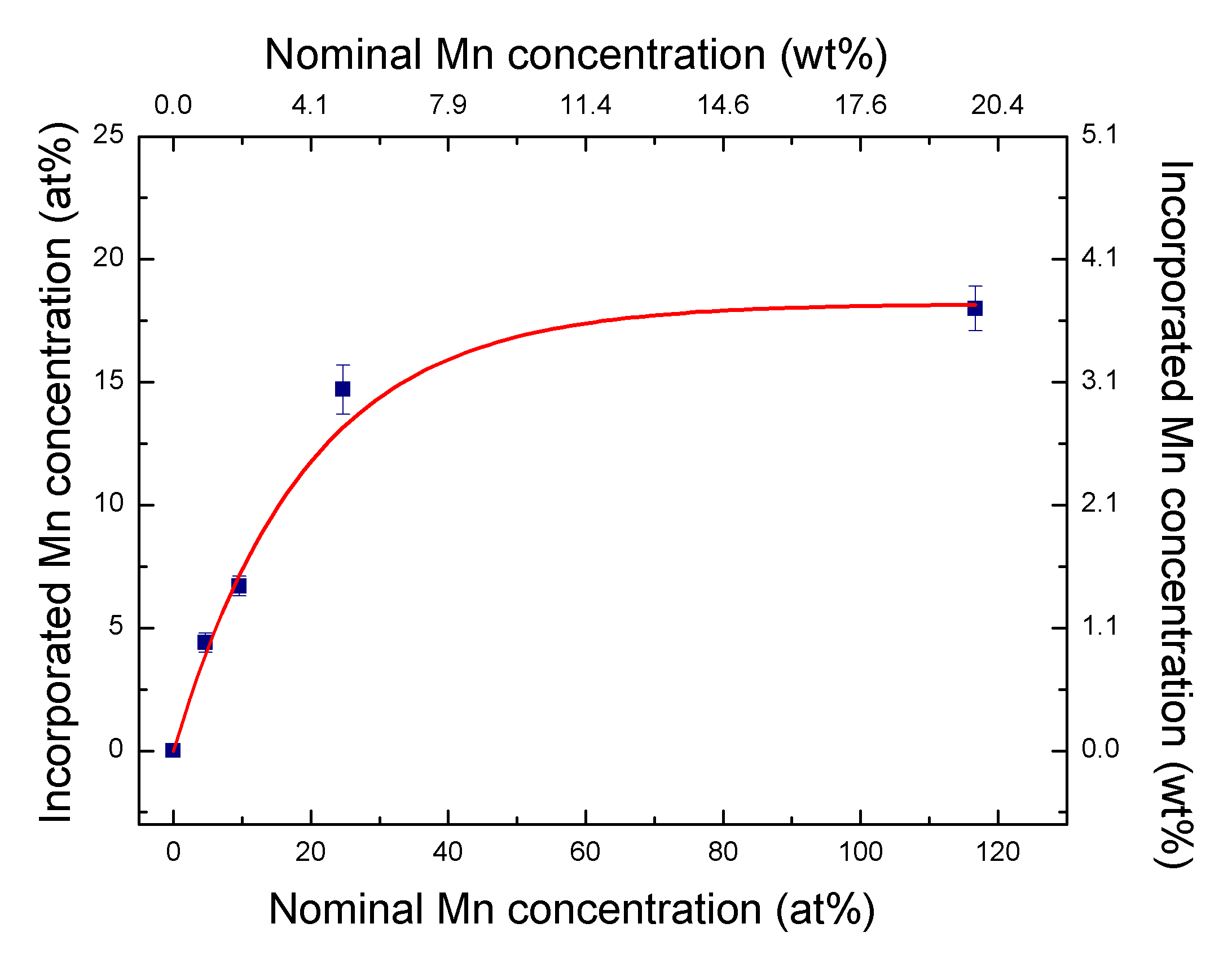}
\caption{Nominal vs. the incorporated Mn concentration ratio for the case of MnH-NTs. Line is a fit to an exponential saturation model $a(1-e^{-bx})$ with $a=18(1)$ at.\%; $b=0.052(6)$. The abbreviations contain the incorporated Mn concentration. The incorporated Mn concentration $x$ of MnH-NT translates to $y=x/(x+3)$ Mn concentration of Mn-NT upon heat treatment.}
\label{edx}
\end{figure}
water and neutralized up to pH$\approx$6.5 with the appropriate amount of 0.1 M HCl solution (Merck). During this step, sodium exchange proceeds to the formation of H$_2$Ti$_3$O$_7$ nanotubes. The sample is finally washed with hot ($80~^{\circ}$C) deionized water in order to remove any traces of NaCl. To dope H$_2$Ti$_3$O$_7$ nanotubes by Mn$^{2+}$, they are suspended in a Mn(NO$_3$)$_2 \cdot$H$_2$O solution at $10~^{\circ}$C for 1~h. The suspension is subsequently filtered and washed with 500~ml of deionized water in order to remove the non-exchanged Mn$^{2+}$ remaining in the solution. As a result, a solid Mn$_x$H$_{2-x}$Ti$_3$O$_7$ phase is obtained with x up to 0.18 (MnH-NTs). For Mn$^{2+}$ doped TiO$_2$ NTs, in the final step MnH-NTs undergo heat treatment at 400 $^{\circ}$C in a reducing atmosphere (N$_2$/H$_2$) in order to prevent Mn$^{2+}$ oxidation into higher oxidation states and a single phase Mn$_y$Ti$_{1-y}$O$_2$ is obtained (Mn-NTs).\\
The manganese content was determined by energy-dispersive X-ray spectroscopy (EDX). XRD measurements were performed in $\theta/2\theta$ geometry on powder samples using Cu K$_\alpha$($\lambda$=1.54056 \AA) radiation. The morphology of the samples was examined by low-/high-resolution transmission electron microscopy (TEM/HRTEM). Electron spin resonance (ESR) measurements of the nanotubes were carried out in an X-band spectrometer in the 5 K to 300 K temperature range. SQUID measurements were performed on a S600 magnetometer following the zero-field-cooled/field-cooled (ZFC/FC) magnetization measurements. 

\section{Results and discussion}
\begin{figure}[tb]
\includegraphics*[width=0.9\linewidth,height=0.73\linewidth]{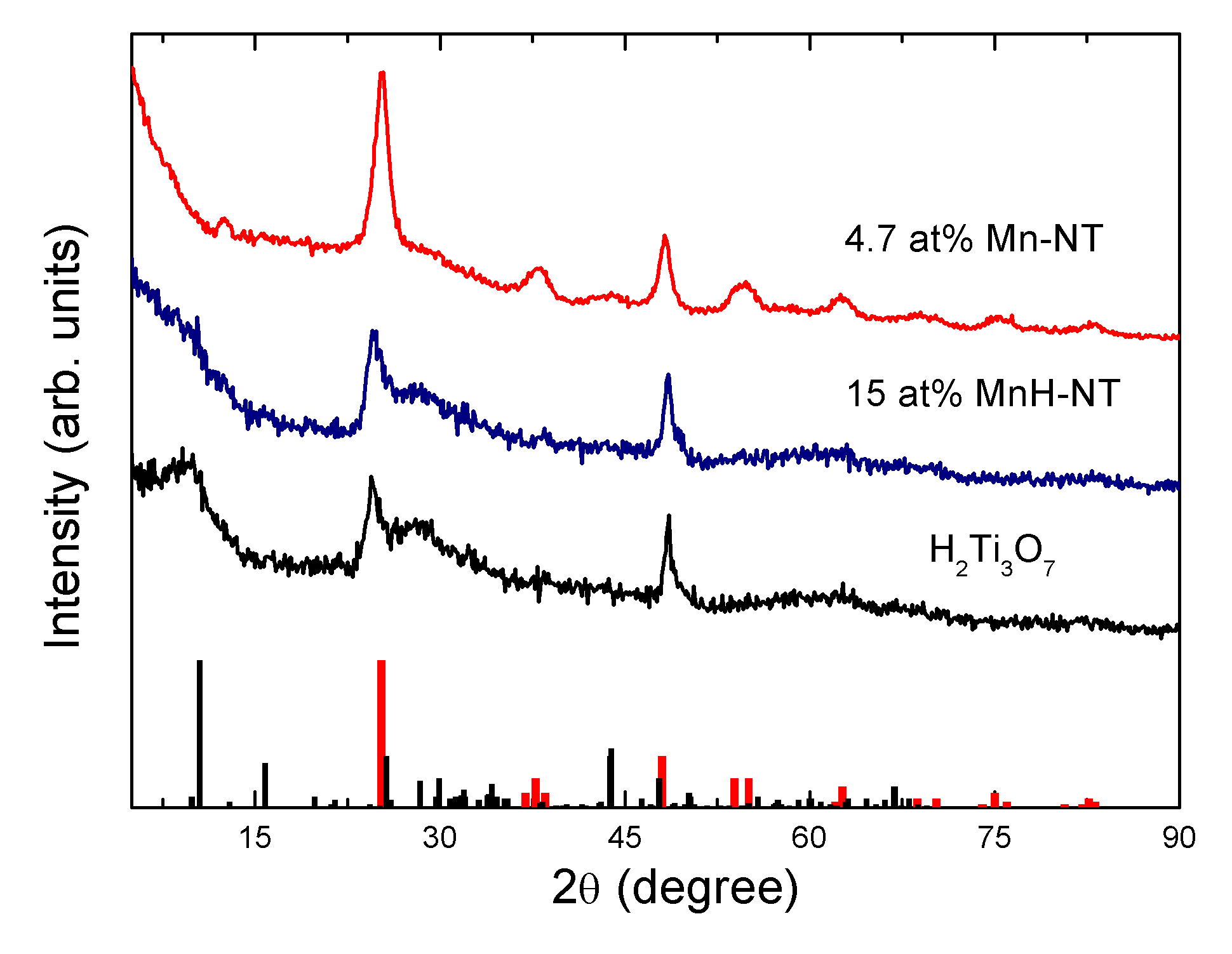}
\caption{XRD spectra of undoped titanate (H$_2$Ti$_3$O$_7$) nanotubes, 15 at.\% Mn doped titanate nanotubes (15 at.\% MnH-NT), and 4.7 at.\% Mn doped TiO$_2$ (4.7 at.\% Mn-NT). The peak at 2$\theta$=13$^{\circ}$ in 4.7 at.\% Mn-NT originates from the sample holder.}

\label{xrd}
\end{figure}
In Fig.~\ref{edx}, the manganese ion concentration of MnH-NTs assuming complete ion-exchange (nominal) versus the equilibrium manganese concentration (incorporated) after the ion-exchange is depicted based on EDX measurements. The line in Fig.~\ref{edx} is a fit to an exponential saturation model $a(1-e^{-bx})$. It yields $a=18(1)$ at.\% saturation concentration, and $b=0.052(6)$ characteristic exchange ratio. This doping level and Mn$^{2+}$ exchange efficiency are seen as characteristics to the described synthesis method.\\
The kinetics of alkali metal ion intercalation between the layers of titanate nanotubes and nanofibers from aqueous suspension has been thoroughly studied by Bavykin \textit{et~al} \cite{bavykin}. They found that the limiting stage of the process is likely to be the diffusion of ions inside the solid crystal which strongly depended on the length of the nanotubes. Here we focus on the elucidation of the state, local interaction and spatial distribution of Mn$^{2+}$ in nanotubular titanates and their derivatives after the steady-state concentration has been reached.\\
The powder XRD data measured on the MnH-NTs and Mn-NTs are given in Fig.~\ref{xrd}. The XRD pattern of the undoped, protonated titanate nanotubes can be indexed as the monoclinic trititanate (H$_2$Ti$_3$O$_7$) phase.\cite{cho2007, sekino} The characteristic reflection near 2$\theta$=10$^{\circ}$ is correlated with the interlayer distance $d_{200}$ in the wall of nanotubes. XRD profile of the 15 at.\% MnH-NT compared to the undoped, protonated sample (H$_2$Ti$_3$O$_7$) shows the weakening of the peak near 2$\theta$= 10$^{\circ}$ upon the ion-exchange. Similar weakening of this characteristic reflection was found by several authors \cite{wang2007, morgado, sunli}. This could be related to the distortion of crystalline order within the layers due to ion-exchange.\\
The heat treatment of MnH-NTs in reduced atmosphere resulted in the creation of Mn-NTs where TiO$_2$ exists in anatase phase TiO$_2$ (JCPDS 84$\underline{\text{ }}$1285). No other peaks of minority phases as manganese titanate, metallic manganese or its oxides have been observed.\\
In principle, in a solid the ion-exchange positions are spatially well defined places. Therefore, we expect that the ionic exchange across the titanate nanoscrolls ensures that manganese ions do not create aggregates in the MnH-NT structure. Furthermore, the relatively low-temperature heat treatment process for making the Mn-NTs suppresses the diffusion and aggregation of Mn ions.\\
\begin{figure}[tb]
\includegraphics*[width=0.9\linewidth,height=0.73\linewidth]{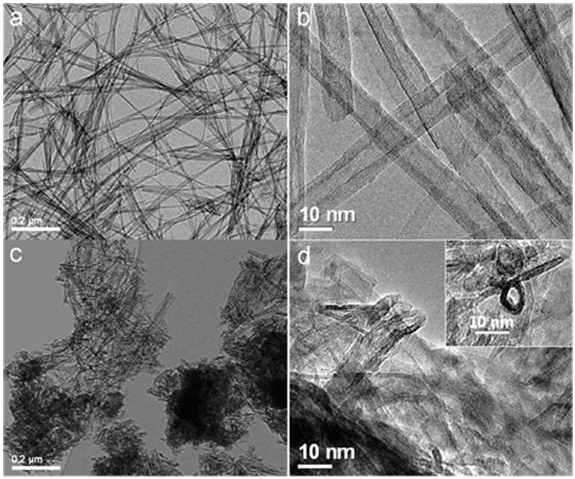}
\caption{(a) TEM and (b) HRTEM image of pristine titanate nanotubes; (c) TEM and (d) HRTEM image of a 4.7 at.\% Mn-NTs produced under heat-treatment of MnH-NTs at $400~^{\circ}\text{C}$ in a reducing atmosphere.}

\label{TEM}
\end{figure}
The TEM images of 4.7 at.\% Mn-NT, produced under heat-treatment at 400 $^{\circ}$C ascertain the prevailing tubular morphology of the doped TiO$_2$ material. Unlike the pristine trititanate nanotubes, the manganese-doped TiO$_2$ nanotubes are crystallized together near the contact point during the heat treatment, which results in highly aggregated nanotube secondary structure (Fig.~\ref{TEM}.c-d). The effect of calcination on the morphological evolution and phase transition of protonated titanate nanotubes has been studied by many researchers. It has been reported that upon calcination the gradual interlayer dehydration leads to a titanate-to-anatase crystal phase change accompanied by the transformation of the tubular shape into nanorods. Electron microscopic investigations revealed that the nanotubes still retain the tubular shape at 350-400 $^{\circ}$C \cite{zhang2007}.\\
\begin{figure}[tb]
\includegraphics*[width=0.9\linewidth,height=1.7\linewidth]{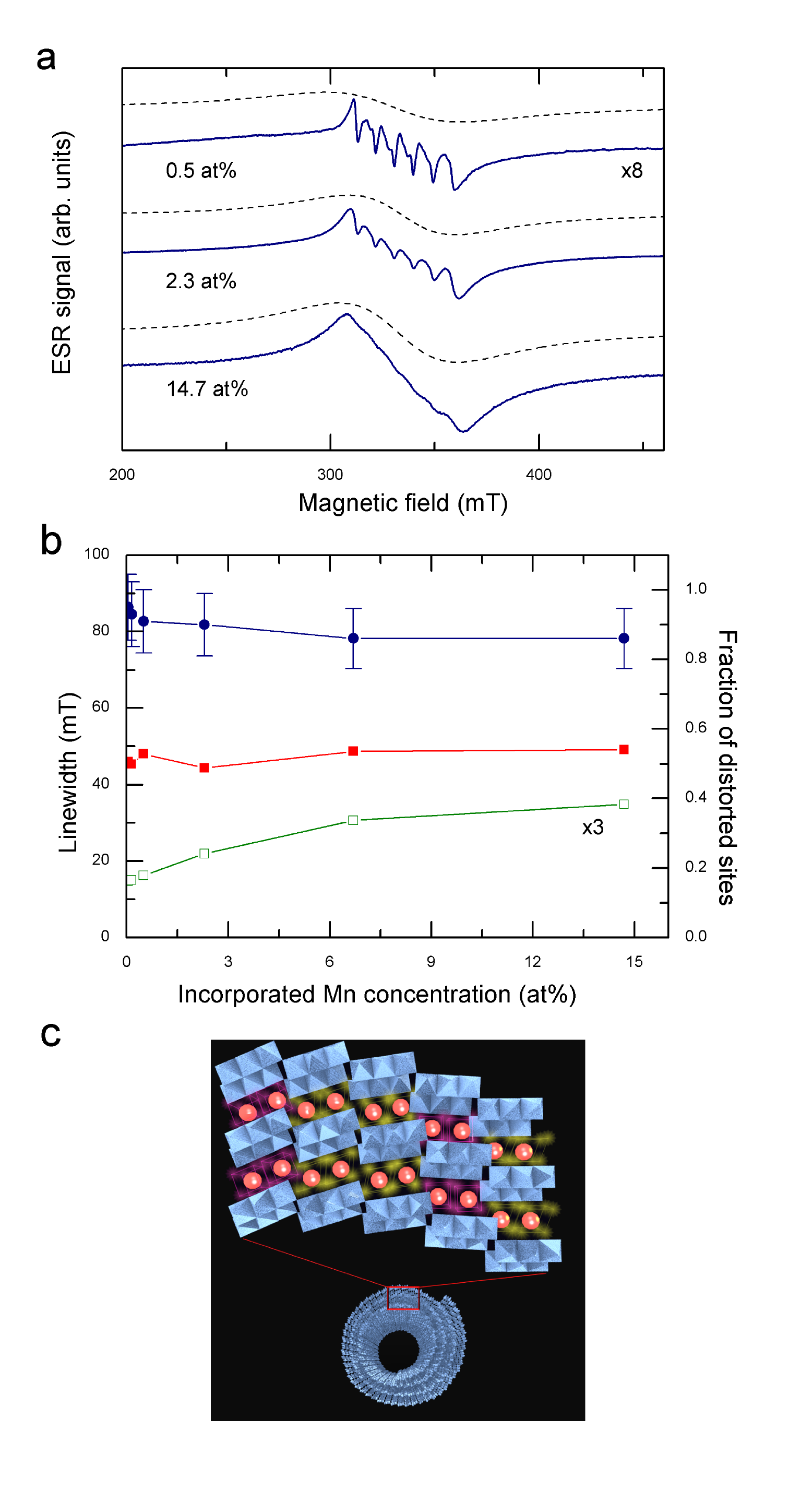}
\caption{(a) Room-temperature ESR measurements on 0.5, 2.3, and 14.7 at.\% MnH-NTs. Spectra are superpositions of two ESR signals: a narrow sextet and a 48 mT broad line. Dashed lines are fits to the broad component.  (b) ESR line width of individual hyperfine lines (\textcolor{green}{$\square$}) and of the broad component (\textcolor{red}{$\blacksquare$}) as a function of Mn concentration. Mn concentration dependence of the fraction of distorted Mn sites (\textcolor{blue}{$\bullet$}). (c) Schematic view of the tubular structure of MnH-NTs. Purple and yellow tetrahedrons represent slightly and strongly distorted Mn sites, respectively.}

\label{NHT}
\end{figure}
\begin{figure*}[tb]

\includegraphics*[width=0.9\linewidth,height=0.36\linewidth]{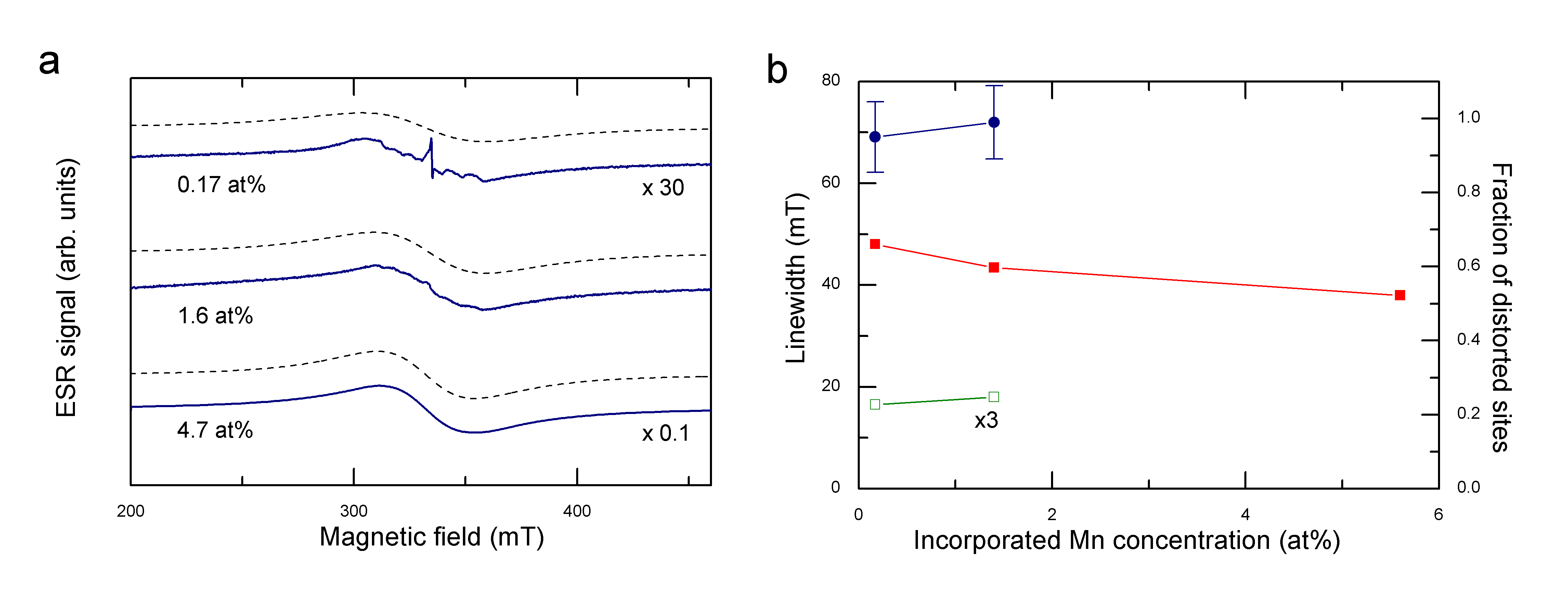}
\caption{Room-temperature ESR measurements on 0.17, 1.6, at.\% and on 1.6 at.\% Mn-NTs. Spectra are superpositions of three ESR signals: a narrow sextet, a 46 mT broad line (dashed lines), and the SETOV. (b) ESR line width of individual hyperfine lines (\textcolor{green}{$\square$}) and of the broad component (\textcolor{red}{$\blacksquare$}) as a function of Mn concentration. Mn concentration dependence of the fraction of distorted Mn sites (\textcolor{blue}{$\bullet$}). At Mn concentration higher than 1.6 at.\%, the strong broadening of the sextet lines impedes the sextet lines to be resolved. This explains the absence of data for the line width and fraction of distorted sites.}

\label{HT}
\end{figure*}
Neither XRD nor TEM measurements find segregated Mn-rich phases in the studied materials. Nevertheless, these techniques are limited to identifying about percent concentration minority phases. ESR, in contrast, is a dopant-specific spectroscopic technique that provides highly detailed microscopic information about the oxidation state and the local environment of the substituent paramagnetic ions in the crystal on ppm levels in principle. Furthermore, ESR is sensitive to the interactions between the paramagnetic ions. ESR spectroscopy has proven to be a very useful method for evaluating success in the synthesis of colloidal Mn$^{2+}$-doped semiconductor nanocrystals \cite{counio, saponjic}. The theoretical background of ESR spectra of a large number of nanocrystals doped with Mn$^{2+}$ has now been largely developed \cite{allen, misra}. The spin Hamiltonian that describes the spectra of Mn$^{2+}$ is
\begin{equation*}
\mathcal{H}=\mathcal{H}_{\text{Z}}+\mathcal{H}_{CF}+\mathcal{H}_{HF}+\mathcal{H}_{(e-e)}
\label{cf}
\end{equation*}
where first is the Zeeman term, second ($\mathcal{H}_{CF}$) is the interaction with crystal electric fields. The third term ($\mathcal{H}_{HF}$) describes the hyperfine coupling between the $S=5/2$ electron spin and $I=5/2$ nuclear spin of Mn$^{2+}$. The last term ($\mathcal{H}_{(e-e)}$) is the interaction between neighboring Mn$^{2+}$ ions.\\
Characteristic ESR spectra of MnH-NTs are given in Fig.~\ref{NHT}.a. They confirm the presence of magnetically isolated Mn$^{2+}$ ions in the titanate nanotube structure. The spectra at all Mn$^{2+}$ concentrations consist of two signals. One set of sextet lines dominates the spectra at low Mn$^{2+}$ concentrations. In contrast, a 48 mT broad line is more pronounced at high Mn$^{2+}$ concentrations. \\
The sextet signal is characteristic of a hyperfine splitting of $^{55}$Mn with $g=2.001(1)$ $g$-factor and $A_{\text{iso}}=9.1$ mT hyperfine coupling constant. This spectrum corresponds to allowed ($\Delta m_I$=0) and forbidden ($\Delta m_I=\pm1$) hyperfine transitions between the Zeeman sublevels $m_s=\pm1/2$. It is characteristic to Mn$^{2+}$ ions in octahedral crystal fields. However, since forbidden transitions are observable, Mn$^{2+}$ ions do not occupy strictly cubic sites, as strictly cubic centers have zero probability of forbidden transitions, but the distortion relative to the cubic symmetry is small. The hyperfine coupling constant $A_{\text{iso}}$ is essentially independent of the Mn$^{2+}$ concentration. The width of the individual hyperfine lines increases by increasing the Mn$^{2+}$ concentration (Fig.\ \ref{NHT}.b). Assuming only the dipole-dipole interaction between Mn$^{2+}$ ions as an origin of the broadening following Ref.\ \cite{vanvleck} yields, $d_\text{Mn-Mn}=1$ nm average distance at 15 at.\%. This is a lower bound for the average Mn-Mn distance.\\
The broad Lorentzian signal around g$\approx$2 (dashed line in Fig.~\ref{NHT}.c) is superimposed to the narrow hyperfine sextet. The intensity of this broad line increases with the initial Mn$^{2+}$ content; thus, it is intrinsic to ion exchange. The 48 mT width of the broad line is essentially independent of Mn$^{2+}$ concentration. It is about 5 times the $A_{\text{iso}}=9.1$ mT which is characteristic of Mn$^{2+}$ ions located in strongly distorted octahedral sites. In powdered polycrystalline systems noncentral transitions ($m_S\neq1/2$) are always broadened and unresolved. This is due to the strong angular dependence of the lines and a parameter distribution in both hyperfine- and fine-coupling parameters. Dipolar interactions further broaden the lines. As previously determined, dipole-dipole interaction between Mn$^{2+}$ ions broadens the sextuplet lines corresponding to the central $\Delta m_I$=1 transition. However, dipole-dipole interaction is not strong enough to smear out the hyperfine structure and thus produce the broad line. These findings are characteristic of Mn$^{2+}$-doped nanocrystals like Mn:CdS and Mn:TiO$_2$ \cite{counio, saponjic}.\\
The width of the broad line and the ratio $I_{\text{broad}}/(I_{\text{broad}}+I_{\text{hyperfine}})=0.85$ are found to be independent of the incorporated Mn$^{2+}$ concentration for MnH-NTs (Fig.~\ref{NHT}.c). Here $I_{\text{broad}}$ is the ESR intensity of the broad line, $I_{\text{hyperfine}}$ is the ESR intensity of the sextuplet. This also confirms that the broad Lorentzian curves in Fig.~\ref{NHT} originate from distorted Mn$^{2+}$ sites. \\
The question to be answered here is: what is the origin of the two different ion-exchange positions in the TiO$_6$ octahedra built host matrix? We speculate that the otherwise defectless layers of TiO$_6$ octahedra are not lined up in perfect registry due to the rolled up morphology of the trititanate nanotubes \cite{kukovecz2005, gateshki}. Along this line, the 15\% doping-independent fraction of high-symmetry Mn$^{2+}$ positions are more likely a consequence of the incommensurate facing of titanium-centered octahedrons in neighboring walls due to the rolled up structure of trititanate nanotubes (Fig.~\ref{NHT}.b). Another possible explanation would be that structural imperfections (i.e., defects) have formed during the oriented crystal growth of the nanotubes, when these TiO$_6$ building bricks were arranged in space \cite{kukovecz2005}.\\
During heat treatment, MnH-NTs undergo dehydration to yield Mn-NTs. During the process, Mn$_x$H$_{2-x}$Ti$_3$O$_7$ is transformed into Mn$_y$Ti$_{1-y}$O$_2$. Mn/Ti stoichiometry remains constant, so that $y=x/(x+3)$. For example, 5.6 at.\% Mn-NTs originates from 18 at.\% MnH-NTs.\\
Fig.~\ref{HT} depicts ESR on 0.17, 1.6 and 4.7 at.\% Mn-NTs obtained after thermal treatment of the corresponding MnH-NTs. The spectra show the coexistence of a sextuplet and a broad Lorentzian line around $g\approx2$ originating from Mn$^{2+}$ ions similarly to Mn-NTs. Furthermore, a narrow symmetric line close to the free electron $g$-factor ($g$=2.003) was also observed. This additional narrow line at $g=$2.003 can be assigned to the single-electron-trapped oxygen vacancies (SETOV) \cite{cho2007, serwicka1981, serwicka1985}. The SETOV signal is induced by breaking of the Ti-OH bonds which follows the dehydration upon heat treatment. This is characteristic to all annealed titania nanotubes \cite{cho2007, counio}. The line width of the SETOV signal is $ \Delta B_{\text{pp}}$=0.5 mT in accordance with the literature \cite{cho2007, cho2009}. As already reported (in the 140 K to 300 K temperature range) \cite{cho2009}, ESR intensity of the SETOV satisfies Curie-law, confirming the presence of localized electrons. ESR measurements on the undoped TiO$_2$ (not shown) reveal the presence of the SETOV as well.  At doping levels lower than 0.17 at.\% (not shown), the SETOV signal dominates the spectra but at all doping levels the SETOV concentration remains in the ppm level. At high Mn concentration, the broad Lorentzian curve overwhelms other signals.\\
The ESR intensity is essentially the same before and after the heat treatment of MnH-NTs. This indicates that Mn ions preserve their 2+ valence state during the heat treatment and Mn-NT formation. Furthermore, it is evidenced that the Mn$^{2+}$ distribution remains homogeneous and no Mn$^{2+}$ clustering occurs during heat treatment. Similarly to MnH-NTs, the broad ESR signal stems from Mn$^{2+}$ ions in distorted sites while the sextuplet at $g$=2.001 proves that some Mn$^{2+}$ ions occupy high symmetry sites \cite{counio, saponjic}. However, due to the strong broadening of the sextet lines by increasing Mn concentration, the sextet lines were only resolved at low Mn concentrations. The hyperfine coupling constant is $A_{\text{iso}}=$12 mT in the case of Mn-NT. The ratio of the two different Mn$^{2+}$ lines is 5:95, which indicates that most of the Mn$^{2+}$ ions occupy distorted sites (Fig.~\ref{HT}.b). The ratio increases relative to the MnH-NTs, as a consequence of the different tube geometry.\\
\begin{figure}[tb]
\includegraphics*[width=0.9\linewidth,height=1.31\linewidth]{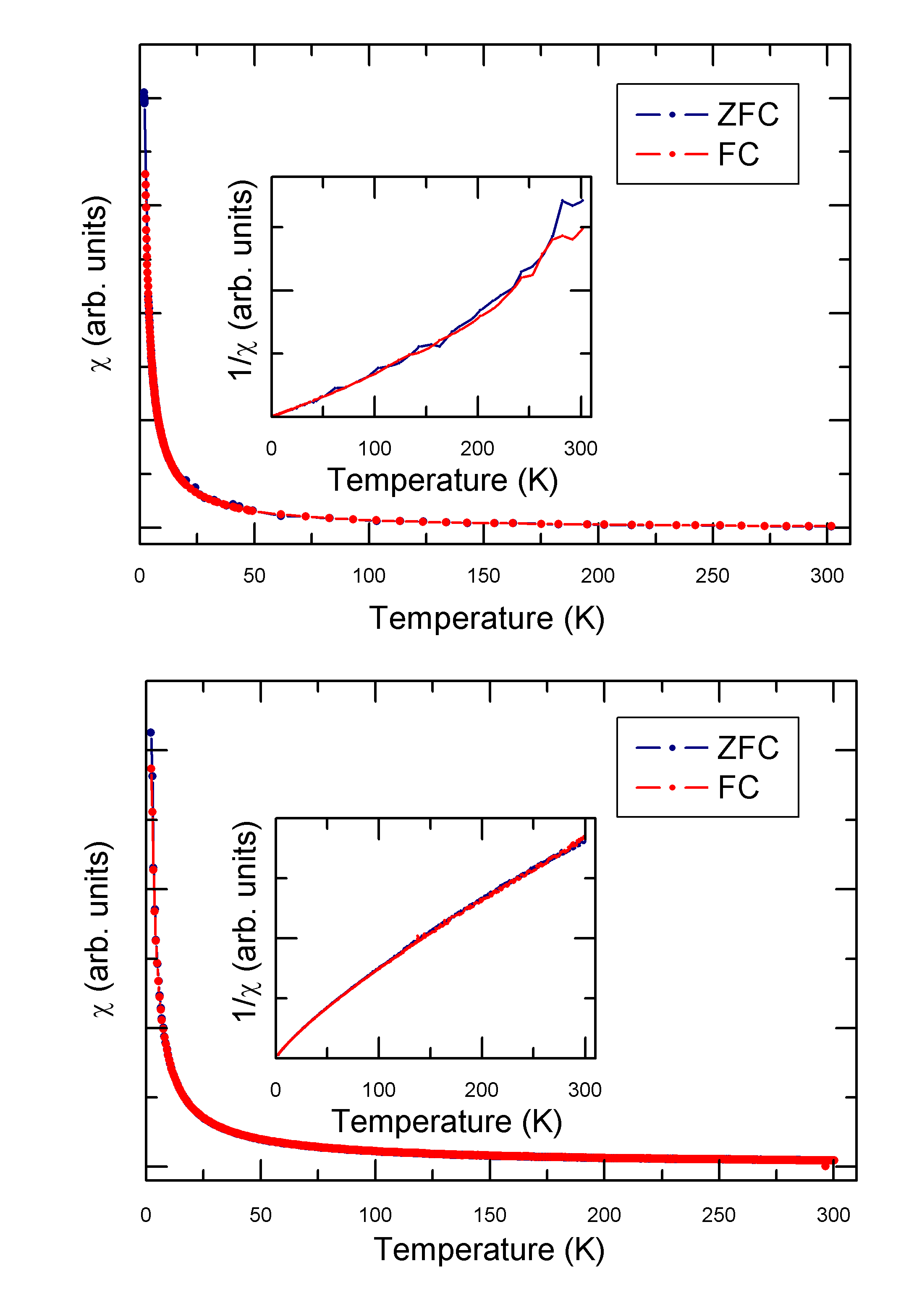}
\caption{ZFC/FC SQUID measurements (a) 0.17 at.\% and (b) for 4.7 at.\% Mn-NT. For FC, the applied field was 0.1 T. In the main panel the susceptibility ($\chi$) is shown as a function of temperature. Inset presents 1/$\chi$ as a function of temperature.}

\label{SQUID}
\end{figure}

ESR found the Mn-MT samples paramagnetic without a trace of magnetically ordered minority phase. However, X-band ESR is not always detectable on magnetically long-range ordered systems; thus, to further characterize the magnetic properties of Mn-NTs, SQUID measurements are required. Fig.~\ref{SQUID} depicts the ZFC/FC SQUID measurements for 0.17 and 4.7 at.\% Mn-NT. The bulk susceptibility is of a Curie-like paramagnetic temperature dependence. This is further emphasized plotting 1/$\chi$ as a function of temperature in the inset of Fig.~\ref{SQUID}. The SQUID susceptibility is in agreement with the ESR intensity of the broad Lorentzian curves assigned to Mn$^{2+}$ ions.\\
Earlier reports suggest that ferromagnetism in the fourth naturally occurring TiO$_2$ polymorph called TiO$_2$(B) is induced by the SETOV \cite{huang2007}. It has been proposed that the ferromagnetic coupling is induced through the F-center exchange mechanism \cite{coeyapl}. Our measurements do not contradict these results. ESR intensity, i.e., the spin-susceptibility of the SETOV, is negligible compared to that of Mn$^{2+}$ ions; thus, the paramagnetic behavior of 0.17 at.\% Mn-NT is provoked by Mn$^{2+}$ ions.

\section{Conclusion}
We have prepared homogeneously Mn$^{2+}$ substituted H$_2$Ti$_3$O$_7$ (MnH-NTs) nanotubes by ion-exchange. Two symmetrically different ion-exchange sites are identified by ESR spectroscopy. About 15\% of the substituted Mn$^{2+}$ ions occupy cubic sites, whereas 85\% shows strongly distorted octahedral local symmetry. Low-temperature heat treatment transforms the titanate structure to Mn$_y$Ti$_{1-y}$O$_2$ nanotubes (Mn-NTs) while it maintains the homogeneity of the Mn$^{2+}$ substitution. We report on the absence of ferromagnetic properties of single phase Mn$_y$Ti$_{1-y}$O$_2$ nanotubes which favors the scenario that ferromagnetism is of extrinsic origin in Mn$_y$Ti$_{1-y}$O$_2$.\\
The synthesis method presented here opens a novel pathway to synthesizing high-quality homogeneously diluted magnetic semiconducting TiO$_2$ nanotube powders. Besides the commonly used doping techniques employing vapor deposition or ion bombardment, the ion-exchange-based protocol could be used as a cost-effective, alternative way to obtain materials that cannot be prepared by heating mixtures of different precursors. Whereas these days vacuum deposition techniques are capable of controlling carrier concentrations moderately well, wet chemical methods to electronic doping are only now being developed and still suffer from many of the above-mentioned challenges. Here we have also demonstrated that ESR-active, ion-exchangeable cations can be seen as local probes to extract so far undetected information about the atomic level structure of nanotubular titanates and their derivatives.\\
The illustrated synthesis method could have long-term importance for the potential use of the ion-exchangable titanate nanotube and nanowire powders in the DMS (Diluted Magnetic Semiconductor) field.

\section{Acknowledgments}
This work was supported by the European project MULTIPLAT, by the Swiss National Science Foundation IZ73Z0$\underline{\text{ }}$128037/1, and by the Swiss Contribution SH/7/2/20. 

\end{document}